\documentclass[aps,twocolumn,showpacs,groupedaddress]{revtex4}
\usepackage{amssymb}
\usepackage{graphicx}

\newcommand{\ket}[1]{|#1\rangle}
\newcommand{\bra}[1]{\langle#1|}

\begin{document}

\title{Suppressing decoherence and improving entanglement by
quantum-jump-based  feedback control in two-level systems}
\author{S. C. Hou}
\author{X. L. Huang}
\author{X. X. Yi}
\affiliation{School of Physics and Optoelectronic Technology,\\
Dalian University of Technology, Dalian 116024 China}

\date{\today}

\begin{abstract}
We study the quantum-jump-based  feedback control on the
entanglement shared between two qubits with one of them subject to
decoherence, while the other qubit is under the control. This
situation is very relevant to a quantum system consisting of nuclear
and electron spins in solid states.  The possibility to prolong the
coherence time of the dissipative qubit is also explored. Numerical
simulations  show that the quantum-jump-based feedback control can
improve the entanglement between the qubits and prolong the
coherence time for the qubit  subject directly to decoherence.
\end{abstract}
\maketitle
\section{introduction}
Superposition of states and  entanglement make quantum information
processing  much different from its classical counterpart. But a
quantum state would unavoidably  interact with its environment,
resulting in a degradation of coherence and entanglement. For
example, spontaneous emission in atomic qubits \cite{Roos} would
spoil  the coherence of quantum states and  limit the entanglement
time.

Recent experimental advances have  enabled individual systems to be
monitored and manipulated at quantum level \cite{Puppe}. This makes
the quantum feedback control realizable. Among the feedback
controls, The homodyne-mediated feedback \cite{Wiseman,Wiseman2} and
quantum-jump-based feedback controls have been  proposed to generate
steady state entanglement  in a cavity \cite{Wang,Carvalho}. These
two feedback schemes are Markovian, namely,  a  feedback information
proportional to the quantum-jump detection is synchronously  used.
Besides, these control scheme can also be used to suppress
decoherence \cite{Viola,Katz,Ganesan,Zhang}.

Meanwhile, researchers are looking for  proper  systems  for
experimental implementation of quantum information processing. Among
the various candidates, solid-states quantum devices based on
superconductors \cite{Bertet} and lateral quantum dots
\cite{Hayashi} are promising ones,  however, the decoherence from
intrinsic noise originating from two-level fluctuators is hard to
engineered \cite{Rebentrost}. For this reason, the nuclear spins have
attracted considerable attention \cite{Vandersypen} due to its  long
coherence times \cite{Ladd}. But their weak interactions to others
make the preparation, control, and detection on them difficult.
Thanks to its intrinsic interactions with electron spins, electron
spin can be used as an ancilla  to access single nuclear spin. This
naturally leads us to rise the following question: can feedback
strategy be used to suppress decoherence, prepare and protect
entanglement between the nuclear and electron spins  by controlling
the electron spin? In this paper, we will study this problem by
considering a nuclear spin (as a qubit) coupled to electron spin (as
the other qubit) that is exposed to its environment. We show that a
Markovian feedback based on quantum-jump can be used to suppress
decoherence, produce entanglement and protect it.

The paper is organized as follows:  In Sec.{\rm II}, we describe our
model and present the dynamics  in absence of feedback. In Sec.{\rm
III}, we introduce  the quantum-jump-based feedback control and give
the dynamical  equation under the feedback control. The effect of
feedback control on decoherence and entanglement is discussed in
Sec.{\rm IV} and Sec.{\rm V}, respectively. Sec.{\rm VI} concludes
our results.

\section{model}
Our system consists of a pair of  two-level systems, called qubit 1
and qubit 2, where only the  qubit 2 interacts with its environment.
We present a scheme employing quantum-jump-based feedback control on
the  qubit 2 to affect  the decoherence of the  qubit 1 and increase
entanglement between the two qubits. The Hamiltonian of the system
reads
\begin{eqnarray}
H=\frac{1}{2}\hbar\omega_1\sigma_{1}^z+\frac{1}{2}\hbar\omega_2\sigma_{2}^z+\hbar
g(\sigma_1^+\sigma_2^-+\sigma_1^-\sigma_2^+) .
\label{eqn:systemhamiltonian}
\end{eqnarray}
The first two terms represent the free  Hamiltonian of the two
qubits, the last term describes their interactions under the
rotating-wave approximation. $\omega_1$ and $\omega_2$ are the
transition frequency of the two qubits, respectively. $g $ is the
coupling strength of the two qubits. $\sigma_z$ is the Pauli matrix,
i.e., $\sigma_z=\ket{e}\bra{e}-\ket{g}\bra{g}$, and
$\sigma^+=\ket{e}\bra{g}$, $\sigma^-=\ket{g}\bra{e}.$

The  state of this quantum system can be  described by the density
operator $\rho$ which is obtained by tracing out the environment.
The dynamics of open quantum systems can be described by quantum
master equations. The most general form of master equation for the
density operator is \cite{quantumoptics,quantumnoise}
\begin{eqnarray}
\dot{\rho}=-\frac{i}{\hbar}[H,\rho]+\mathcal{L}(\rho),
\label{eqn:masterequation}
\end{eqnarray}
where $H$ is the system Hamiltonian  and $\mathcal{L}$ is a
superoperator defined by $\mathcal{L}(\rho)=\Sigma_k\gamma_k(L_k\rho
L_k^{\dag}-\frac{1}{2}L_k^\dag L_k\rho-\frac{1}{2}\rho L_k^\dag
L_k),$ in which different $k$ characterizes  different  dissipative
channels.

In our system, the first qubit is assumed  to be isolated from
environment. The decoherence comes from  the spontaneous emission of
the qubit 2 (the second qubit). This situation is of relevance to a
system consisting of nuclear and electron spins in aforementioned
solid state devices. The dynamics of such a system takes,
\begin{eqnarray}
\dot{\rho}=-\frac{i}{\hbar}[H,\rho]+\gamma(\sigma_2^-\rho\sigma_2^+-
\frac{1}{2}\sigma_2^+\sigma_2^-\rho-\frac{1}{2}\rho
\sigma_2^+\sigma_2^-) .
\label{eqn:full}
\end{eqnarray}
Here $\sigma_2^{\pm}=I_1\otimes\sigma_2^{\pm}$. The  second part of
Eq.(\ref{eqn:full}) describes the dissipation of our system with
$\gamma$ the decay rate.

Though the first qubit is assumed to be isolated  from environment,
it still loss coherence due to the coupling to the second qubit. The
decoherence process can be showed by the decay of off-diagonal
elements of the reduced density matrix for the first qubit.

In order to investigate this decoherence, we  calculate the
evolution of system density operator $\rho$ and then trace out the
second qubit to get the reduced matrix
\begin{eqnarray}
\rho_1=\text{Tr}_2(\rho)=\sum_{k=e,g}{ }_2\langle k|\rho|k\rangle_2=
\left(
\begin{array}{cc}
\rho_{ee} & \rho_{eg}\\
\rho_{ge} & \rho_{gg}\\
\end{array}
\right).
\end{eqnarray}
The diagonal  elements are the populations in the excited and ground
states of the first qubit. And the off-diagonal elements represent the
coherence of the qubit 1.

\section{Quantum-jump-based Feedback control}
Quantum feedback controls   play an increasingly  important role in
quantum information processing. It is widely used to create and
stabilize  entanglement as well as  combat with decoherence
\cite{Carvalho,Wang,Zhang,Katz}.  In our model, the second qubit is
used as an ancilla  through which the feedback can affect the
dynamics of the first qubit, i.e., by employing a feedback control
on the second qubit, we control the first qubit. The  goal is to
suppress the decoherence of the first qubit and enhance the
entanglement between the two qubits by a feedback control on the
second qubit\cite{Carvalho}.

Our feedback control strategy is based on  quantum-jump detection.
The master equation with feedback can be derived from the general
measurement theory \cite{Wiseman2}. In our paper,
Eq.(\ref{eqn:full}) is equivalent to
\begin{eqnarray}
\rho(t+dt)=\sum_{\alpha=0,1}\Omega_{\alpha}(T)\rho(t)\Omega_{\alpha}^{\dag}(T).
\label{measure}
\end{eqnarray}
with
\begin{eqnarray}
\Omega_{1}(dt)=\sqrt{\gamma dt}\sigma_2^-
\\\Omega_{0}=1-(\frac{i}{\hbar}H+\frac{1}{2}\gamma\sigma_2^+\sigma_2^-)dt.
\label{measure2}
\end{eqnarray}
When the measurement result is $\alpha=1$, a detection occurs, which
causes a finite evolution in the system via $\Omega_1(dt)$. This is
 called a quantum jump. Then the unnormalized density matrix becomes
$\tilde{\rho}_{\alpha=1}=\sigma_2^-\rho(t)\sigma_2^+dt$. The
feedback control is added by giving $\tilde{\rho}_{\alpha=1}$ a finite
unitary evolution, then $\tilde{\rho}_{\alpha=1}$ become
$\tilde{\rho}_{\alpha=1}=F\sigma_2\rho(t)\sigma_2^+F^{\dag}dt$.
In the limit that the feedback acts immediately after a detection and
in a very shot time (much smaller than the time scale of the
system's evolution), the master equation is Markovian,
\begin{eqnarray}
\dot{\rho}=-\frac{i}{\hbar}[H,\rho]+\gamma(F\sigma_2^-\rho\sigma_2^+F^\dag-
\frac{1}{2}\sigma_2^+\sigma_2^-\rho-\frac{1}{2}\rho
\sigma_2^+\sigma_2^-). \label{eqn:controlled}
\end{eqnarray}
Here $F=e^{iH_f}$ and $H_f=-\frac{1}{\hbar}H_f't_f.$  We see that
the operator $H_f$ contains a relatively large operator $H_f'$
multiplied by a very short time $t_f$ (Markovian assumption), but
the product represents a certain amount of evolution, so it is
convenient to discuss $H_f$ instead of $H_f'$ and $t_f$. Here $H_f$
is a $2\times 2$ hermit operator which can be decomposed by Pauli
matrixes $H_f=A_x\sigma_x+A_y\sigma_y+A_z\sigma_z$ ($A_x,A_y,A_z$
are real numbers). So we have,
\begin{eqnarray}
F=I_1\otimes e^{i\vec{A}\cdot\vec{\sigma}}=I_1\otimes(\cos|\vec{A}|+i\frac{\sin|\vec{A}|}{|\vec{A}|}
\vec{A}\cdot\vec{\sigma}).
\label{eqn:feedback}
\end{eqnarray}
Here $\vec{\sigma}=(\sigma_{x},\sigma_{y},\sigma_{z})$ and
$\vec{A}=(A_x,A_y,A_z)$ representing the amplitude of
$\sigma_x,\sigma_y$ and $\sigma_z$ control.

In order to understand  the physical meaning of feedback operator
$F$, we rewrite it as $F=I_1\otimes
e^{-i\frac{\omega}{2}\vec{n}\cdot\vec{\sigma}}$ where
$\vec{n}=(\sin{\theta}\cos{\phi},sin{\theta}\sin{\phi},
\cos{\theta})$ and $\vec{\sigma}=(\sigma_x,\sigma_y,\sigma_z)$, this
feedback operator is equivalent to a time-evolution with evolution
operator $F=I_1\otimes e^{iH_f}$. And it is clear that the operator
$F$ rotate the Bloch  vector of the second qubit with the angle
$\omega$ around the $\vec{n}$ axis. The relationship between the two
forms of $F$ are $A_x=-\frac{\omega}{2}\sin{\theta}\cos{\phi},
A_y=-\frac{\omega}{2}\sin{\theta}\sin{\phi},
A_z=-\frac{\omega}{2}\cos{\theta}.$ So a $\sigma_x$ control
($A_y=0,A_z=0$) means rotating the Bloch vector with a certain
amount of angle around the $x$ axis of Bloch sphere, so does the
$A_y$ and $A_z$ control. Different $\vec{A}$ represents different
feedback evolution i.e., rotate the Bloch vector with a particular
angle around a particular direction in the Bloch sphere. For
simplicity, we discuss the $\sigma_x, \sigma_y,\sigma_z$ control one
by one in the following.

This control mechanism has the advantage of being simple to apply in
practice, since it does not need real time state estimation as
Bayesian feedback control does\cite{Wiseman3}. The emission of the
second qubit is measured by a photo detector, whose signal provides
the information to design the control $F$. In this kind of
monitoring, the absence of signal predominates the dynamics  and the
control is triggered only after a detection click, i.e. a quantum
jump, occurs.

\section{Decoherence suppression}
\begin{figure}
\includegraphics*[width=6cm]{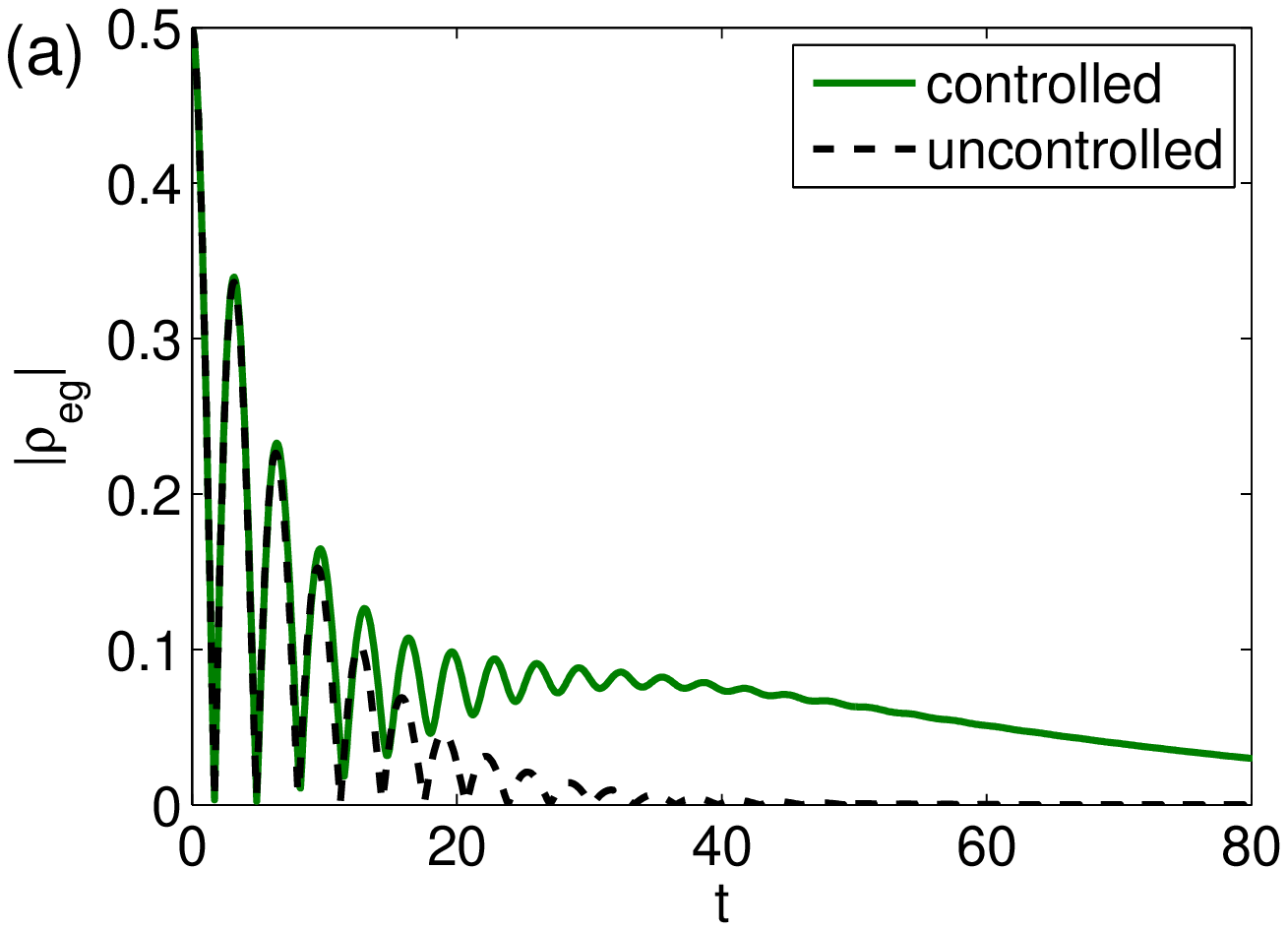}
\includegraphics*[width=6cm]{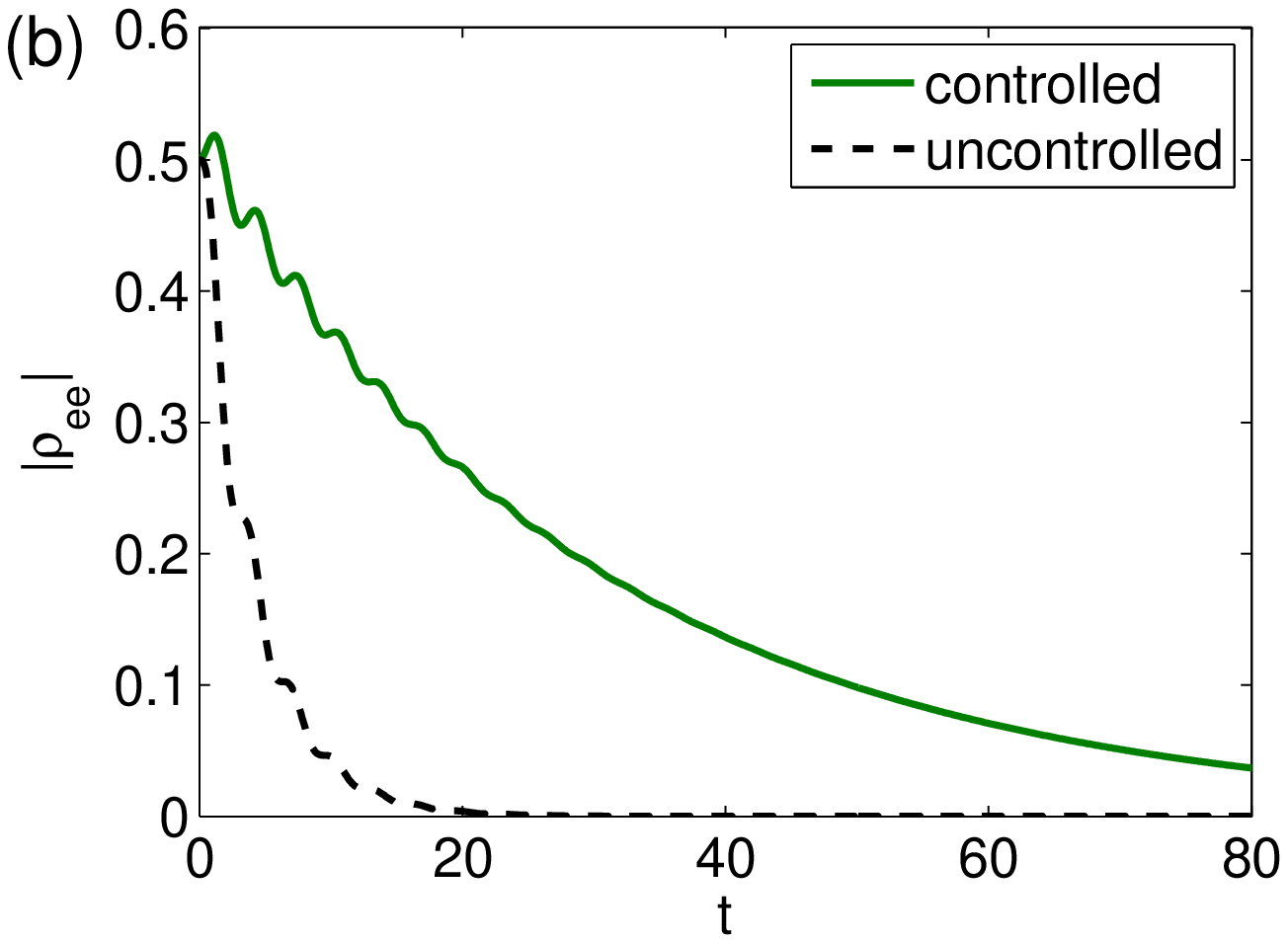}
\caption{(a) Time evolution of  $|\rho_{eg}|$  with and without
control. $t$ is in the unit of $\frac{1}{\omega}$. The different
curves correspond to $A_x=1.2,A_y=A_z=0$(solid line) and
$A_x=A_y=A_z=0=0$ (dashed line), for $g/\omega=1,\gamma/\omega=0.5$.
The feedback control strategy results in an improvement in
decoherence time evidently.  (b) Excited state population
$\rho_{ee}$ evolution with and without control for the same
parameters with (a), the decay of excited state population is slower
in the controlled scheme.} \label{FIG:rhovs}
\end{figure}

Before investigating the influence of the feedback  control, we
first analyze the evolution of our system without control.  Assume
that the two qubits are initially in the same pure superposition
state, for example,
$|\psi\rangle=\frac{1}{\sqrt{2}}(|e\rangle_1+|g\rangle_1)
\otimes\frac{1}{\sqrt{2}}(|e\rangle_2+|g\rangle_2)$. The
corresponding density matrix is,
\begin{eqnarray}
\rho_0=|\psi\rangle\langle\psi|=\frac{1}{4}
\left(
\begin{array}{cccc}
1 & 1 & 1 & 1 \\
1 & 1 & 1 & 1 \\
1 & 1 & 1 & 1 \\
1 & 1 & 1 & 1 \\
\end{array}
\right).
\end{eqnarray}
We assign the Planck constant $\hbar$ to  be 1,
$\omega_1=\omega_2=\omega$ in Eq.(\ref{eqn:systemhamiltonian}), and
$g/\omega=1,\gamma/\omega=0.5$. After numerical calculation, we get
the evolution of the density matrix for the first qubit  without
control.  Since $\rho_{eg}=\rho_{ge}^*, \rho_{ee}+\rho_{gg}=1$, we
only discuss coherence $|\rho_{eg}|$ and excited state population
$\rho_{ee}$ for simplicity. The evolution of  $|\rho_{eg}|$ and
$\rho_{ee}$ without control is depicted in Fig.\ref{FIG:rhovs} (a)
and (b) (dashed lines).

In Fig.\ref{FIG:rhovs} (a), a fast decay  of $|\rho_{eg}|$ (dashed
line)  can be found. This demonstrates that the first qubit lost
coherence due to the second qubit's spontaneous emission and their
interaction. Meanwhile, the first qubit lost energy due to couplings
to the second qubit (Fig.\ref{FIG:rhovs} (b) (dashed line)). The
results also show that the populations in excited state and ground
state decay away. This is  because the first qubit exchange energy
with the second qubit, see Eq.(\ref{eqn:systemhamiltonian}).

Now we add feedback control $F$ to our system,  the master equation
then becomes Eq.(\ref{eqn:controlled}). Our system is initially in
the state $\rho_0$, other parameters remain  unchanged.

\begin{figure}
\includegraphics*[width=6cm]{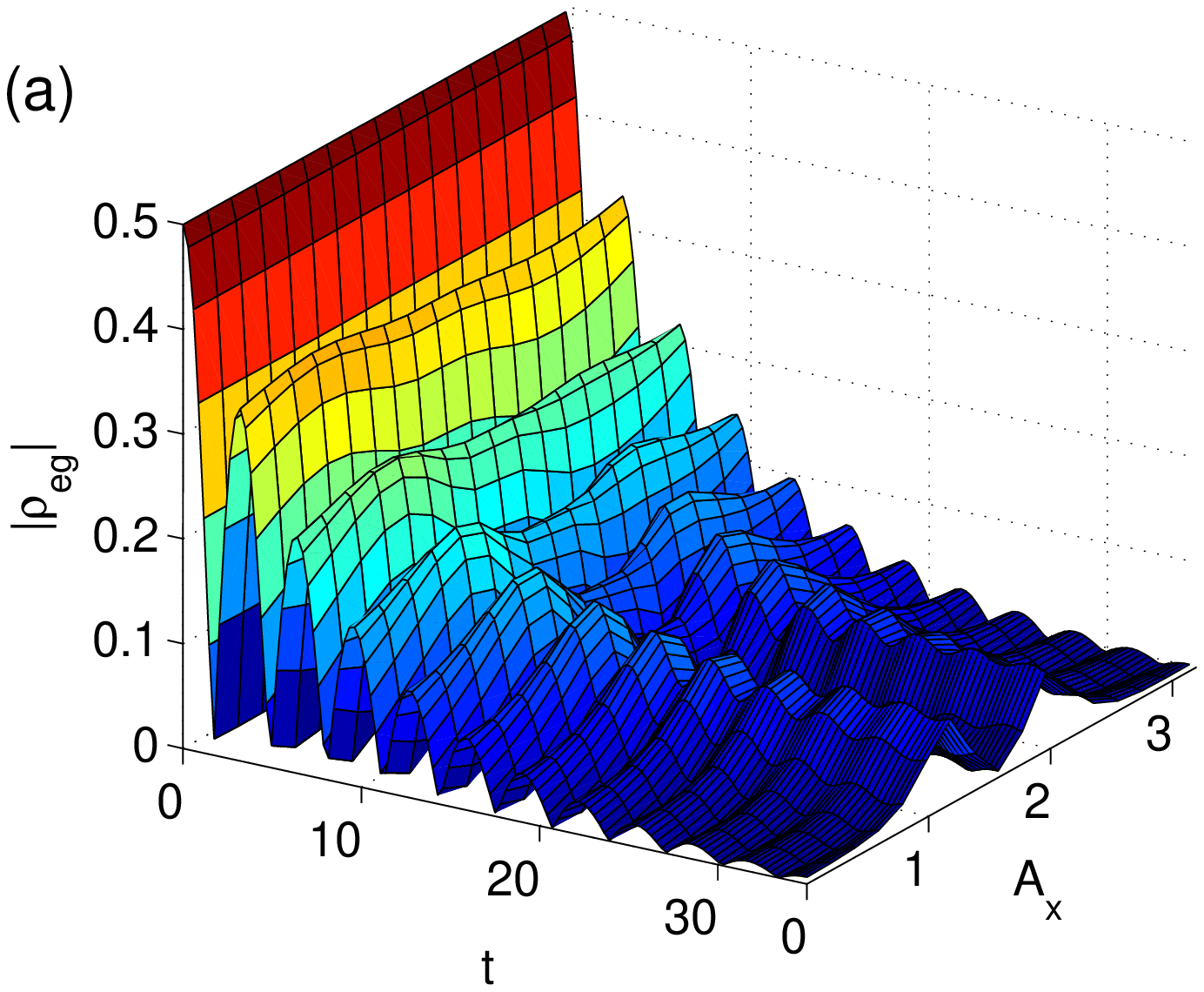}
\includegraphics*[width=6cm]{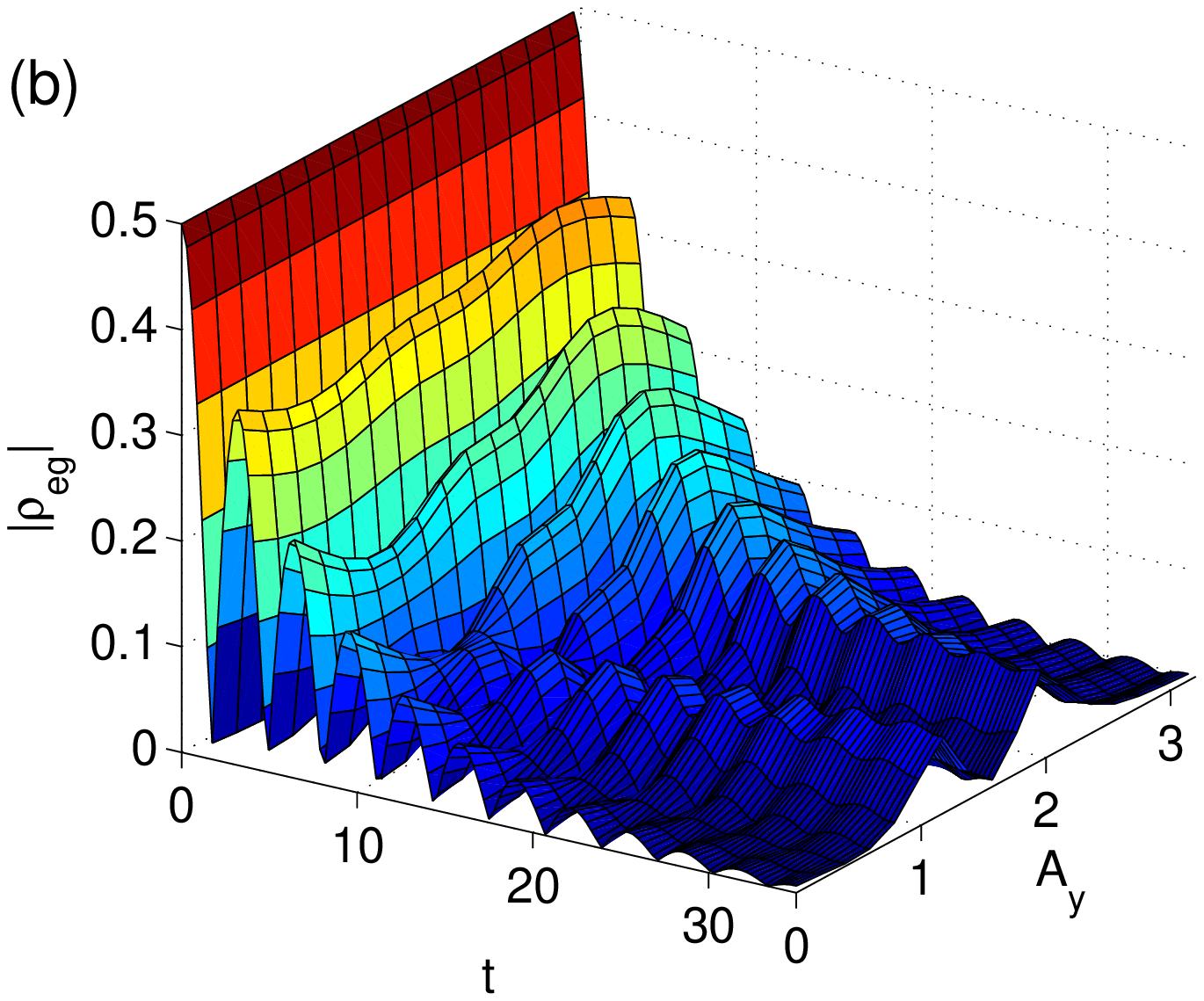}
\includegraphics*[width=6cm]{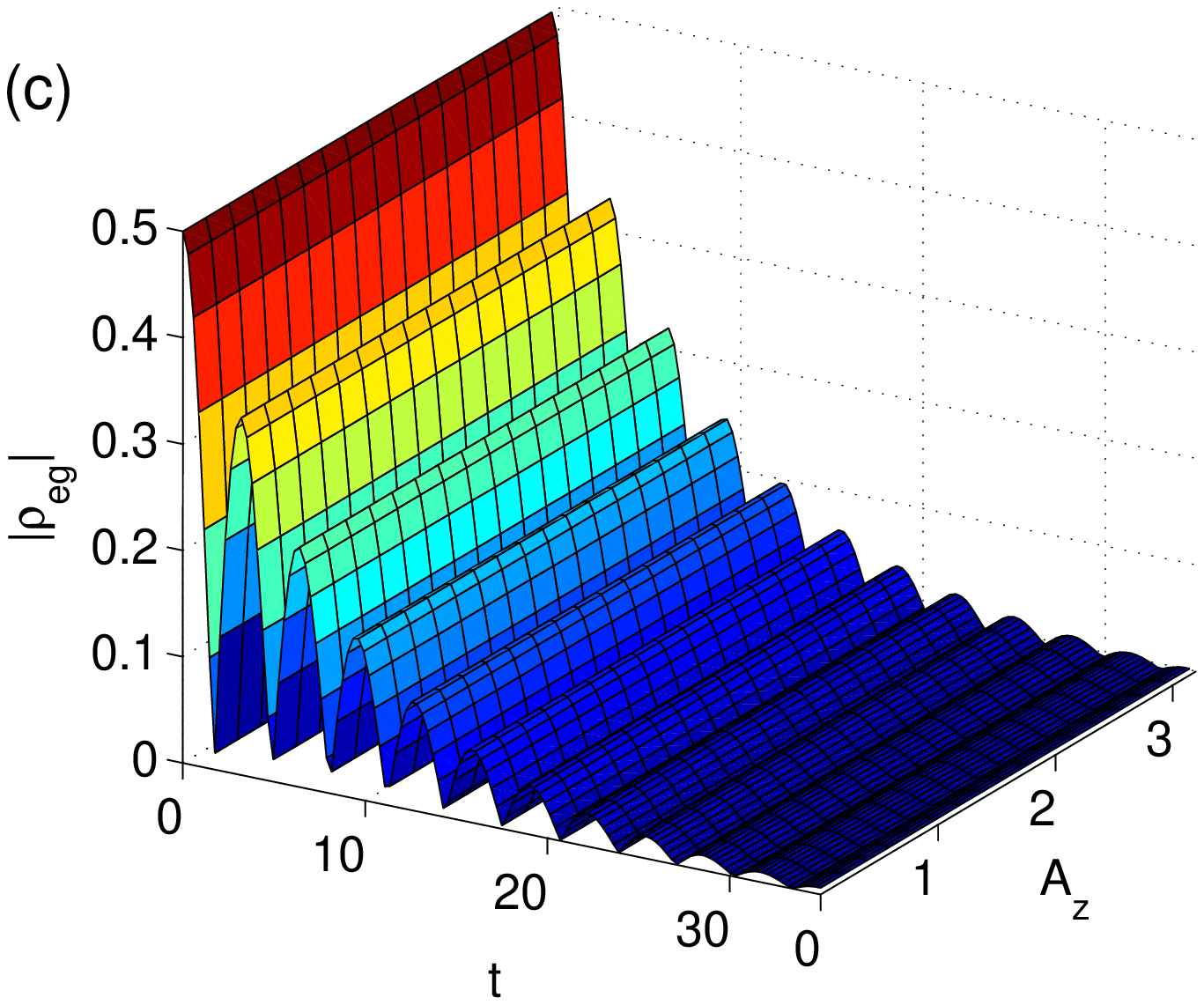}
\caption{The evolution of absolute value of the  first qubit's
off-diagonal element with different control parameters, for
$g/\omega=1,\gamma/\omega=0.5$ and $t$ in the unit of
$\frac{1}{\omega}$. (a) The $\sigma_x$ control for
$A_x=0\thicksim\pi,A_y=A_z=0$. (b) The $\sigma_y$ control for
$A_y=0\thicksim\pi,A_x=A_z=0$. (c) The $\sigma_z$ control for
$A_z=0\thicksim\pi,A_x=A_y=0$. When the feedback amplitude is chosen
to be about 1.3 and 1.9 for both $\sigma_x$ and $\sigma_y$ control,
the oscillation of off-diagonal element is remarkably enhanced. The
$\sigma_z$ control doesn't work in our model. } \label{FIG:rho3}
\end{figure}

We first analysis the $\sigma_x$ control by  choosing feedback
amplitude $A_x=0\thicksim\pi, A_y=A_z=0$. Note that when
$A_y=A_z=0,$ the feedback amplitude $A_x$ influence the system's
evolution with a period of $\pi$ which comes from the term
$F\sigma_2^-\rho\sigma_2^+F^{\dag}$ in Eq.(\ref{eqn:feedback}). It
can be analytically proved that
$e^{iA_x\sigma_x}\sigma^-\rho_2\sigma^+
e^{-iA_x\sigma_x}=e^{i(A_x+\pi)
\sigma_x}\sigma^-\rho_2\sigma^+e^{-i(A_x+\pi)\sigma_x}$ and
$e^{iA_y\sigma_y}\sigma^-\rho_2\sigma^+e^{iA_y\sigma_y}=
e^{i(A_y+\pi)\sigma_y}\sigma^-\rho_2\sigma^+e^{i(A_y+\pi)\sigma_y}$
under any $A_x$ and $A_y$. Here $\rho_2$ is the reduced density
matrix of the second qubit.  The absolute value for the first
qubit's off-diagonal density matrix element evolves as showed in
Fig.\ref{FIG:rho3} (a). The figure indicates that, for an
appropriate feedback amplitude , $A_x\approx1.3$ and
$A_x\approx1.9$, the absolute value of off-diagonal element can be
evidently enhanced compared with the uncontrolled case ($A_x=0$).
That means the decoherence  is partially suppressed. The improvement
of coherence caused by feedback is shown explicitly in
Fig.\ref{FIG:rhovs} (a). We plot $|\rho_{eg}|$, representing the
coherence of the first qubit, as a function of time with
$A_x=1.2,A_y=A_z=0$ (a selected controlled case). In comparison with
the uncontrolled case, a stronger oscillation amplitude and longer
dechoherence time appears. Meanwhile, the $\rho_{ee}$ decays slowly
compared to the uncontrolled case as  shown in Fig.\ref{FIG:rho3}
(b).

Similarly, the $\sigma_y$ control is also able to  slow down the
decay of $|\rho_{eg}|$. We make $A_y=0\thicksim\pi, A_x=A_z=0$. The
numerical results of $|\rho_{eg}|$ is shown in Fig.\ref{FIG:rho3}
(b). Unlike the $\sigma_x$ and $\sigma_y$ control, the $\sigma_z$
control ($A_z=0\thicksim\pi, A_x=A_y=0$) has no effect on the
evolution of the system as shown in Fig.\ref{FIG:rho3} (c). This is
because $e^{iA_z\sigma_z}\sigma_-\rho_2\sigma_+
e^{-iA_z\sigma_z}=\rho_2$ for any $A_z$. The physics behind this
result is that after emitting a photon, the controlled qubit must
stay in the ground state with the Bloch  vector pointing   the
bottom of the Bloch sphere, so the rotation around $z$ axis does not
change the Bloch vector,  i.e., the state of the qubit remains
unchanged.

The present results show that decoherence of  the first qubit can be
suppressed by controlling its partner. The decoherence source in our
system is the spontaneous emission of the second qubit, once the
detector detects  a photon, i.e. a quantum jump of the second qubit
happens, the feedback beam  instantaneously act on the second qubit
and then the first qubit is impacted through the coupling of the two
qubit. The feedback control scheme can reduce the destructive
effects of coherence and slow down the dissipation of energy.

The control effect is relevant to the  coupling strength $g$. When
$g$ is small, the first qubit is hard to be impacted by the second
qubit, so it's hard to prepare, measure and  control the state of
the first qubit. As the interaction goes stronger, the effect of
feedback control becomes more evident.

\begin{figure}
\includegraphics*[width=7cm]{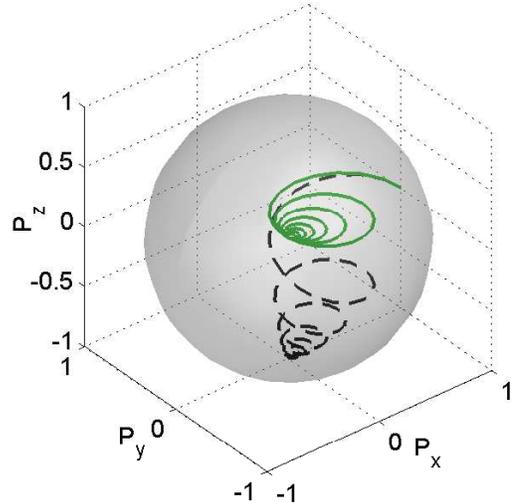}
\caption{ Polarization vector evolution in a  bloch sphere for
feedback amplitude $A_x=\frac{\pi}{2},A_x=A_y=0$ (solid line) and
$A_y=A_x=A_y=0$ (dashed line). The parameters are
$g/\omega=1,\gamma/\omega=0.5$, the initial state is
$|\psi\rangle=\frac{1}{\sqrt{2}}(|e\rangle_1+|g\rangle_1)\otimes\frac{1}
{\sqrt{2}}(|e\rangle_2+|g\rangle_2)$.} \label{FIG:bloch}
\end{figure}

For the case discussed in Fig.\ref{FIG:rhovs}, the first qubit is
dissipative. We found that when the control parameters is chosen as:
$A_x=\frac{\pi}{2},A_y=A_z=0$ , or $A_y=\frac{\pi}{2},A_x=A_y=0$
with the two qubits initially being prepared in the same states, the
decoherence dynamics turns to a phase damping type. The population
in ground state and excited state do not change, while the
off-diagonal elements evolves in  the same way as in the
uncontrolled case. We show this in a bloch sphere\cite{Altafini} in
Fig.\ref{FIG:bloch}. Here the reduced density matrix of the first
qubit can be written by
$\rho_1=\frac{1}{2}(I+\vec{P}\cdot\vec{\sigma})$. We can get the
polarization vector components $P_x=\text{Tr}(\sigma_x\rho_1)$,
$P_y=\text{Tr}(\sigma_y\rho_1)$ and $P_z=\text{Tr}(\sigma_z\rho_1)$.

\section{Entanglement control}
\begin{figure}
\includegraphics*[width=6cm]{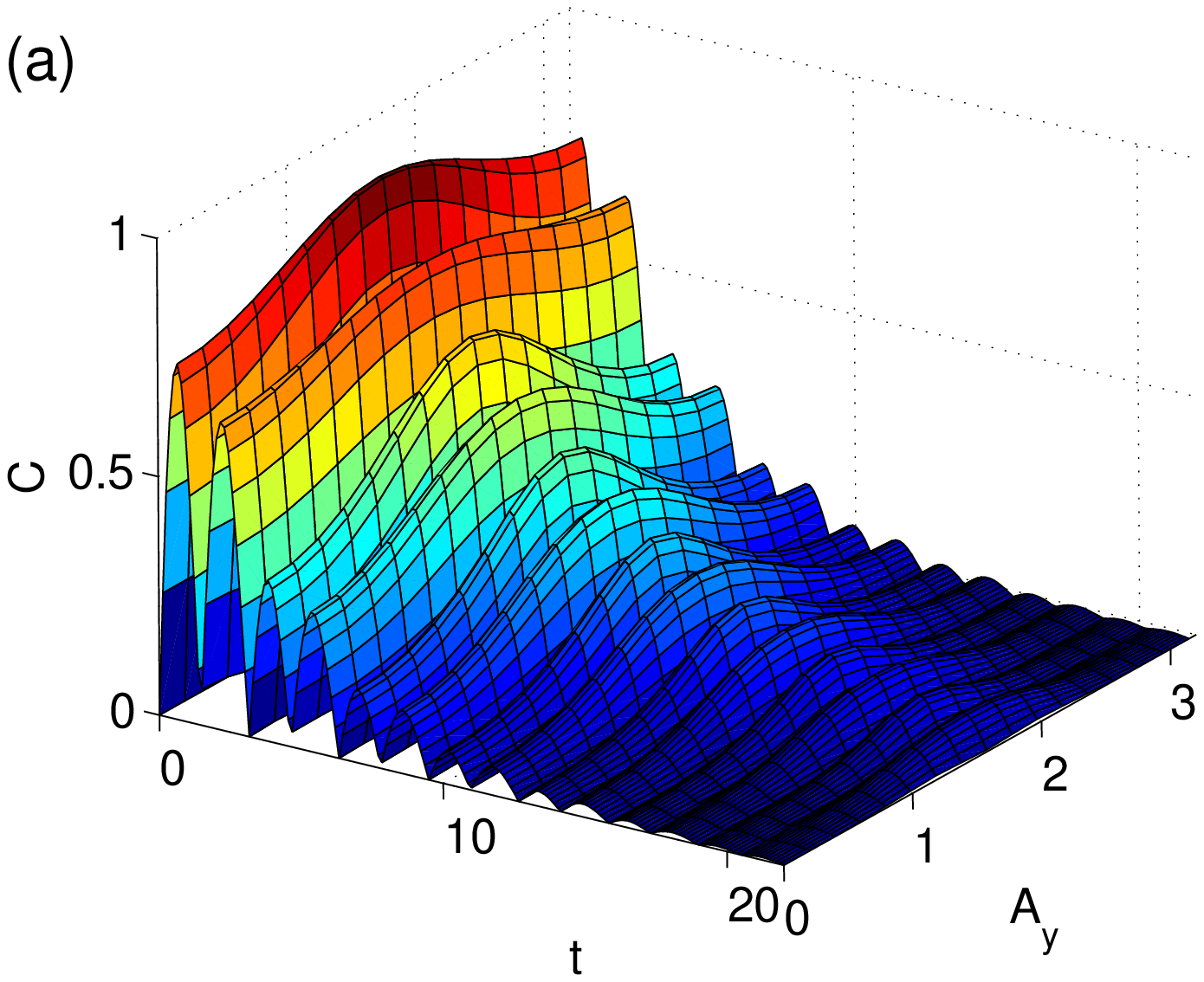}
\includegraphics*[width=6cm]{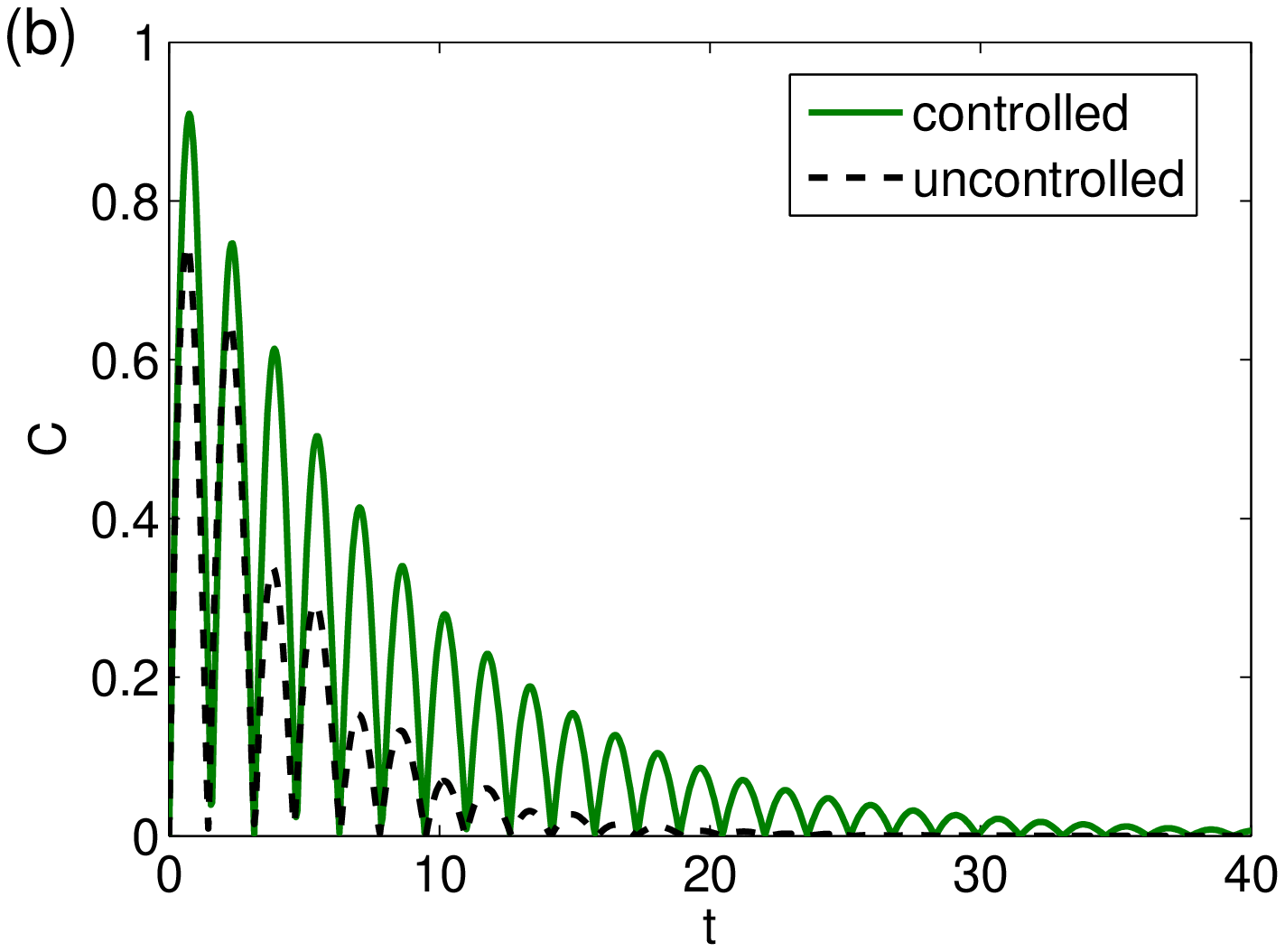}
\caption{(a) Conccurence as a function of  time and $A_y$. The
system is initially in the state
$|\psi\rangle=|g\rangle_1|e\rangle_2$, for the parameters
$g/\omega=1,\gamma/\omega=0.5$. (b) A controlled evolution for
$A_y=0.5\pi,A_x=A_z=0$ vs. uncontrolled case. The entanglement is
improved by choosing an appropriate feedback. $t$ is in the unit of
$\frac{1}{\omega}$ for (a) and (b).} \label{FIG:cge}
\end{figure}

Quantum feedback control has been recently  used to improve the
creation of steady state entanglement in open quantum systems. A
highly entangled states of two qubits in a cavity can be produced
with an appropriate selection of the feedback Hamiltonian and
detection strategy \cite{Carvalho,Carvalho2}. We will show that the
quantum-jump-based feedback scheme can produce and improve
entanglement in our model. We choose the concurrence \cite{Wootters}
as a measure of entanglement. For a mixed state represented by the
density matrix $\rho$ the "spin-flipped" density operator reads
\begin{eqnarray}
\tilde{\rho}=(\sigma_y\otimes\sigma_y)\rho^*(\sigma_y\otimes\sigma_y)
\end{eqnarray}
where the $*$  denotes complex conjugate of  $\rho$ in the bases of
$\{|gg\rangle,  |ge\rangle,  |eg\rangle,  |ee\rangle\}$, and
$\sigma_y$ is the usual Pauli matrix. The concurrence of the density
matrix $\rho$ is defined as
\begin{eqnarray}
C(\rho)=\max{(\sqrt{\lambda_1}-\sqrt{\lambda_2}
-\sqrt{\lambda_3}-\sqrt{\lambda_4},0)}.
\end{eqnarray}
where $\lambda_i$ are eigenvalues of matrix  $\rho\tilde{\rho}$ and
sorted in decreasing order
$\lambda_1>\lambda_2>\lambda_3>\lambda_4$. The range of concurrence
is from 0 to 1, and $C=1$ represents the maximum entanglement.

In absence of spontaneous emission, i.e.  $\gamma=0$, the system
evolves without dissipation. We find that for the system initially
in a separable state except $|\psi\rangle=|e\rangle_1|e\rangle_2$ or
$|\psi\rangle=|g\rangle_1|g\rangle_2$ (the eigenstates of system
Hamiltonian $H$), an entangled state can be generated due to the
interaction between the two qubits. The amount of entanglement
depends on initial states of the system and the coupling strength
$g$. But when the spontaneous emission effect is taken into account,
the performance of  entanglement  preparation   get worse
considerably.

\begin{figure}
\includegraphics*[width=6cm]{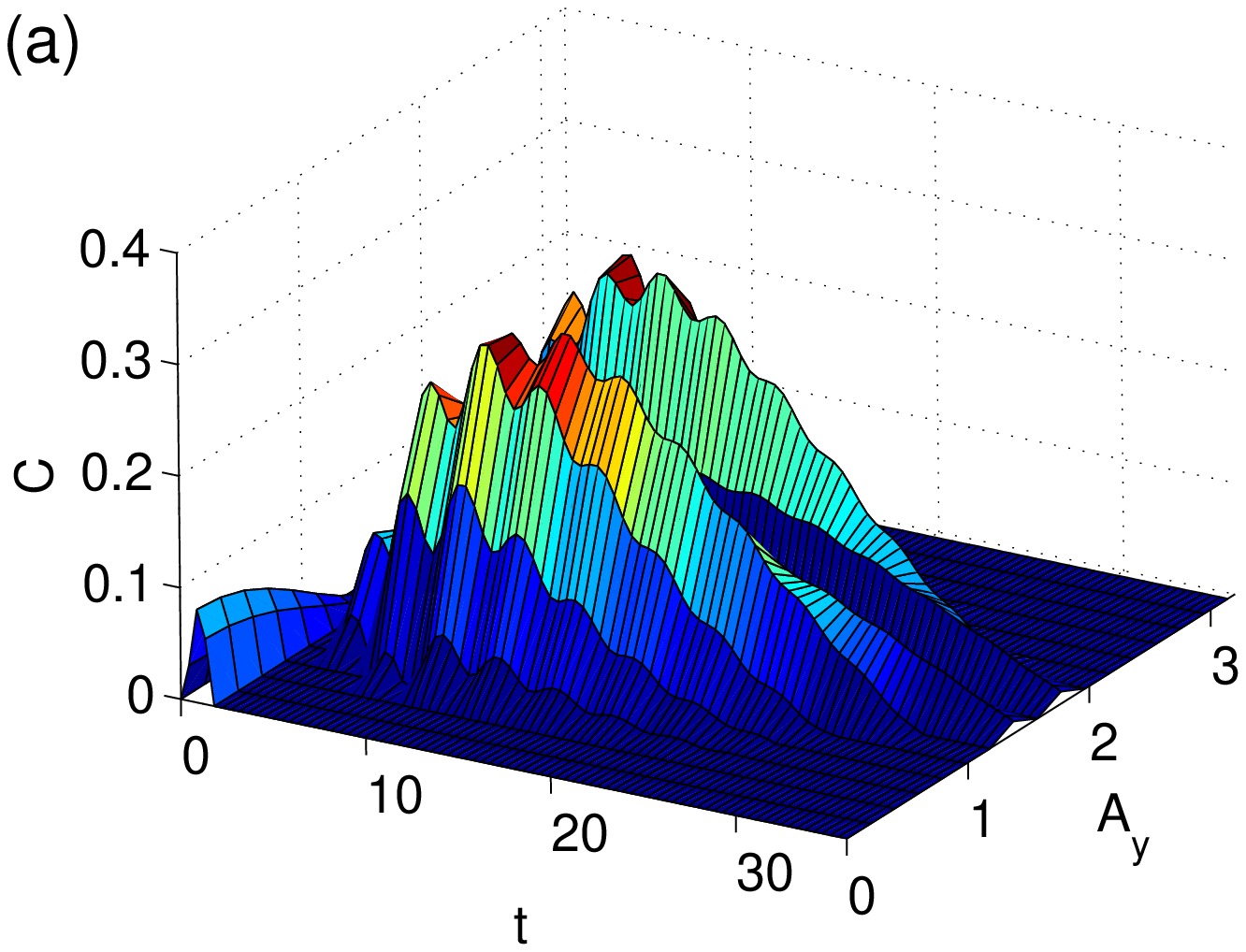}
\includegraphics*[width=6cm]{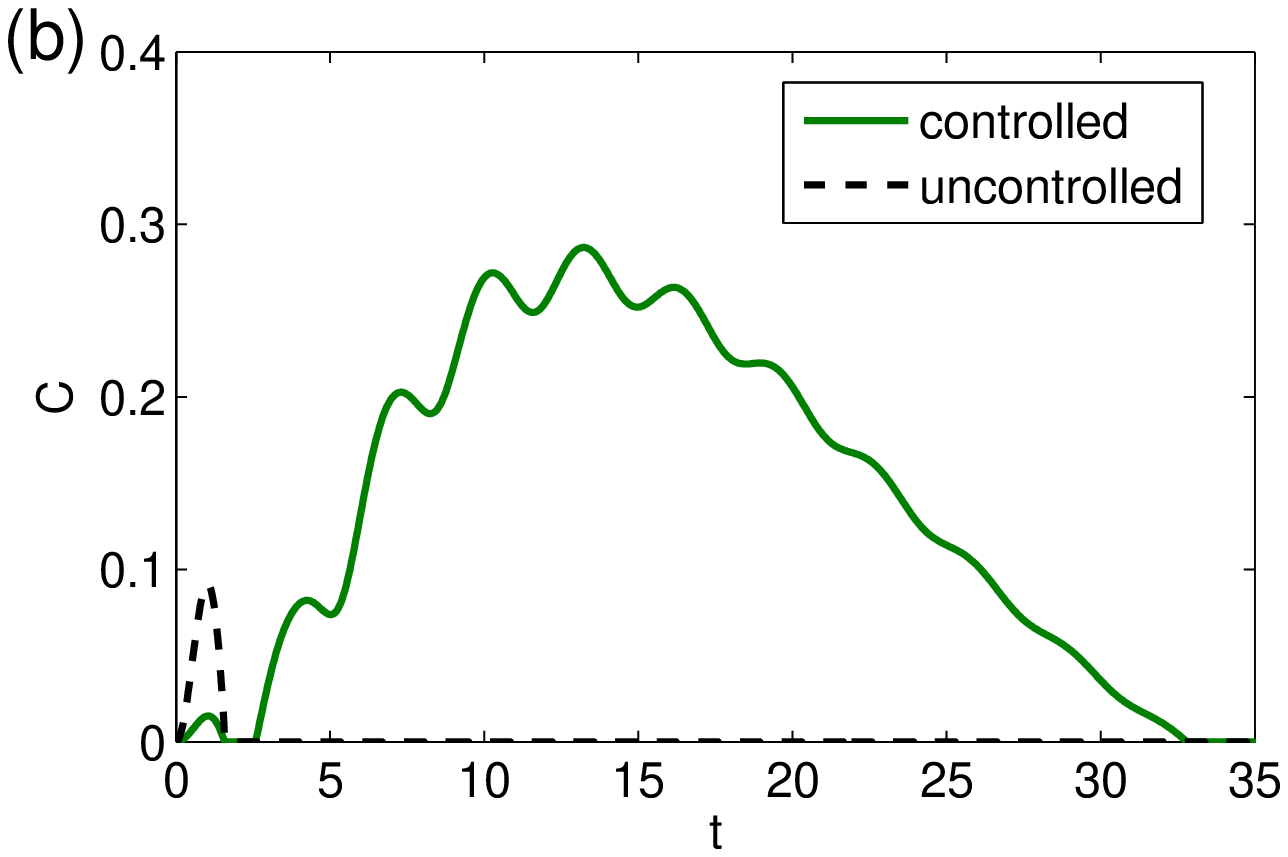}
\caption{ (a) Conccurence as a function of  time and $A_y$. The
system is initially in the state
$|\psi\rangle=|e\rangle_1|e\rangle_2$, for the parameters
$g/\omega=1,\gamma/\omega=0.5$. (b) The controlled conccurence
evolution for $A_y=1.2,A_x=A_z=0$ vs. uncontrolled case. $t$ is
in the unit of $\frac{1}{\omega}$ for (a) and (b).} \label{FIG:cee}
\end{figure}

Now we investigate if our feedback control  strategy can improve the
entanglement preparation  with the effect of spontaneous emission of
the second qubit. The master equation with control is
Eq.(\ref{eqn:controlled}). The effect of feedback control lies in
different choices for the feedback parameters $A_x, A_y, A_z$, the
coupling strength $g$ and different initial states. Here we present
two typical results with two different states.

Our first choice is the initial state
$|\psi\rangle=|g\rangle_1|e\rangle_2$ with $\sigma_y$ control for
$A_y=0\thicksim\pi, A_x=0, A_z=0$. The concurrence evolution is
plotted as a function of time and feedback amplitude $A_y$ in
Fig.\ref{FIG:cge} (a), and Fig.\ref{FIG:cge} (b) denotes the
concurrence evolution with a selected feedback amplitude compared
with the uncontrolled case. We see that entangled  states can be
generated with any feedback parameters, but it decreases with time
because of the dissipative effect. When an appropriate feedback
amplitude  $A_y\approx0.9$ is chosen, the concurrence amplitude is
remarkably enhanced, and the entanglement lasts for a long time. For
the system initially in the state
$|\psi\rangle=|e\rangle_1|e\rangle_2$ with $\sigma_y$ control, the
dynamics of the concurrence  is shown in Fig.\ref{FIG:cee} (a). Note
that in this case if there is no spontaneous effect, this is a
steady state of the system, the density matrix elements does not
change with time. Fig.\ref{FIG:cee} (a) demonstrates that the
dissipation and feedback can produce entanglement. We show this
explicitly in Fig.\ref{FIG:cee} (b) by choosing feedback amplitude
$A_y=1.2$. We can see that for a proper feedback amplitude, after an
entanglement death, a larger amount entanglement is regenerated.

The above results shows the feedback  control strategy can be used
to prepare and protect entanglement in our model. The effect of
entanglement control strongly depends on the initial state. For a
certain initial state, we found that the $\sigma_x$ control and
$\sigma_y$ control has the similar effect but the $\sigma_z$ control
does not work.

\section{Conclusion and remarks}
In this paper, we studied the effect  of quantum jump based feedback
control on a system  consisting  of two qubits  where only one of
them subject to decoherence. By numerical simulation, we found that
it is possible to suppress  decoherence of the first qubits  by a
local control on the second qubits. We observed that  the
decoherence time of the first qubit  is increased remarkably. The
control scheme can also used to protect  the entanglement between
the two qubits. These features can be understood as that the
feedback control changes the dissipative dynamics of the system
through the quantum-jump operators. We would like to note that
Hamiltonian Eq.(\ref{eqn:systemhamiltonian}) does not describe the
hyperfine interaction. However, by the recent technology we can
simulate Hamiltonian Eq.(\ref{eqn:systemhamiltonian}) in
nuclear-electron spin systems, in this sense, the scheme presented
here is available for nuclear-electron spin systems. On the other
hand, by using the hyperfine interaction Hamiltonian, our further
simulations show that we can obtain results similar to that for
Hamiltonian Eq.(\ref{eqn:systemhamiltonian}).


\begin{references}
\bibitem{Roos} C. F. Roos, G. P. T. Lancaster, M. Riebe,
H. H\"{a}ffner, W.H\"{a}sel, S. Gulde, C. Becher, J. Eschner, F.
Schmidt-Kaler, and R. Blatt, Phys. Rev. Lett. \textbf{92}, 220402
(2004).

\bibitem{Puppe} T. Puppe, I. Schuster, A. Grothe,
A. Kubanek, K. Murr, P. W. H. Pinkse, and G. Rempe, Phys. Rev. Lett.
\textbf{99}, 013002 (2007).

\bibitem{Wiseman} H. M. Wiseman,
and G. J. Milburm, Phys. Rev. Lett. \textbf{70}, 548 (1993).

\bibitem{Wiseman2} H. M. Wiseman, Phys. Rev. A \textbf{49}, 2133
(1994).

\bibitem{Wang} J. Wang, H. M. Wiseman, and G. J. Milburn,
Phys. Rev. A \textbf{71}, 042309 (2005).

\bibitem{Carvalho} A. R. R. Carvalho, J.J. Hope,
Phys. Rev. A \textbf{76}, 010301 (2007).

\bibitem{Viola} L. Viola, and S. Lloyd,
Phys. Rev. A \textbf{58}, 2733 (1998).

\bibitem{Katz} G. Katz, M. A. Ratner, and R. Kosloff,
Phys. Rev. Lett. \textbf{98}, 203006 (2007).

\bibitem{Ganesan} N. Ganesan, and T. Tarn,
Phys. Rev. A \textbf{75}, 032323 (2007).

\bibitem{Zhang} J. Zhang, C. Li, R. Wu,
T. Tarn, and X. Liu, J. Phys. A \textbf{38}, 6587-6601 (2005).

\bibitem{Bertet} P. Bertet, I. Chiorescu,
G. Burkard, K. Semba, C. J. P. M. Harmans, D. P. DiVincenzo,
and J. E. Mooij, Phys. Rev. Lett. \textbf{95}, 257002 (2005).

\bibitem{Hayashi} T. Hayashi, T. Fujisawa, H. D. Cheong,
Y. H. Jeong, and Y. Hirayama, Phys. Rev. Lett. \textbf{91}, 226804 (2003).

\bibitem{Rebentrost} P.Rebentrost, I.Serban,
T.Schulte-Herbr\"{u}ggen, and F. K. Wilhelm,
Phys. Rev. Lett. \textbf{102}, 090401 (2009).

\bibitem{Vandersypen} L. Vandersypen and I. Chuang,
Rev. Mod. Phys. \textbf{76}, 1037 (2004).

\bibitem{Ladd} T. Ladd, D. Maryenko, Y. Yamamoto, E. Abe, and K. Itoh,
Phys. Rev. B \textbf{71}, 014401 (2005).

\bibitem{quantumoptics} M. O. Scully, and
M. S. Zubairy {\it Quantum Optics} (Cambridge University Press, Cambridge 1997).

\bibitem{quantumnoise} C. W. Gardiner, and P. Zoller,
{\it Quantum Noises} (Springer-Verlag, Berlin 1991).

\bibitem{Wiseman3} H. M. Wiseman, S. Mancini, and
J. Wang, Phys. Rev. A \textbf{66}, 013807 (2002).

\bibitem{Altafini} C. Altafini, J. Math. Phys. \textbf{44}, 2357 (2003).

\bibitem{Carvalho2} A. R. R. Carvalho, A. J. S. Reid,
and J. J. Hope, Phys. Rev. A \textbf{78}, 012334 (2008).

\bibitem{Wootters}W. K. Wootters, Phys. Rev. Lett. \textbf{80}, 2245 (1998).

\end{references}
\end{document}